\documentclass[12pt]{iopart}
\usepackage{graphicx}% Include figure files
\usepackage{iopams}

\usepackage{bm}% bold math

\begin{document}
\title [Superconducting SmFeAsO$_{1-x}$F$_{x}$]{Evidence of structural phase transition in superconducting SmFeAsO$_{1-x}$F$_{x}$ from $^{19}$F NMR}
\author{M. Majumder\dag, K. Ghoshray\dag\footnote[1]{To whom correspondence should be addressed
(kajal.ghoshray@saha.ac.in)}, C. Mazumdar\dag\, A. Poddar\dag\, A. Ghoshray\dag\, D. Berardan\ddag, N. Dragoe\ddag}

\address{\dag\ ECMP Division, Saha Institute of Nuclear Physics, 1/AF Bidhannagar,
Kolkata-700064, India}
\address{\ddag\ Institut de Chimie Mol\'{e}culaire et des Mat\'{e}riaux d'Orsay, Univ. Paris-Sud 11, UMR 8182, 91405 Orsay, France}

\date{\today}
\begin{abstract}

We report the resistivity, magnetization and $^{19}$F NMR results in polycrystalline sample of SmFeAsO$_{0.86}$F$_{0.14}$. The resistivity and the magnetization data show a sharp drop at 48 K indicating the superconducting transition. The nuclear spin-lattice rate (1/$T_1$) and spin-spin relaxation rate (1/$T_2$) clearly show the existence of a structural phase transition near 163 K in the sample which also undergo superconducting transition. This finding creates interest to explore whether this is unique for only Sm based  systems or it is also present in other rare-earth based 1111 superconductors.

\end{abstract}
 \ead{kajal.ghoshray@saha.ac.in}
\maketitle

\section{Introduction}
The existence of competition between structural, magnetic and superconducting (SC) transition \cite{Johnston10} in F doped oxypnictides makes them unconventional. Such competing order parameters would lead to a complex phase diagram including several coexisting phases in the nano scale range. The general belief is that, the parent compound \textit{Re}FeAsO (\textit{Re}= rare-earth) shows a structural phase transition (SPT) from tetragonal (T) to orthorhombic (O) symmetry around $T_s$=160 K and then nearly 20 K below $T_s$, there is a long range magnetic order (MO) with spin density wave (SDW) type transition. As F doping is increased, transition temperatures for SPT and MO gradually decrease and at a particular concentration of F, the system undergoes SC transition \cite{Margadonna09,Liu08}.

Based on synchrotron powder x-ray diffraction and muon spin rotation studies, a new phase diagram for SmFeAsO$_{1-x}$F$_x$ system has been proposed indicating that the SPT (T - O) in the range 175-155 K survives up to optimal F doped sample \cite{Martinelli11}. The difficulty to determine the occurrence of the SPT is due to the decrease of O distortion with the increase of F content. SPT can be magnetically driven, relieving magnetic frustration resulting from AFM neighbor and next neighbor interactions between local Fe moments \cite{Xu08,Yildirim08}. It can also be related to the nematic ordering \cite{Fang08} defined as the spontaneously broken C4 rotational symmetry of the square lattice due to either AFM spin-fluctuations or orbital ordering as was found in cuprates \cite{Hinkov08} and in the 122 family \cite{Chu10}. Theoretical results also show that in 1111 family, the SPT is driven by spin-fluctuations or orbital ordering \cite{Krüger09}. Martinelli \etal \cite{Martinelli11} claimed that the survival of SPT even for the optimal doped superconducting sample is due to the orbital ordering mediated symmetry breaking force and not driven by magnetic fluctuations.

 We intend to study SmFeAsO$_{0.86}$F$_{0.14}$ using nuclear magnetic resonance (NMR) which is a very useful microscopic tool to probe the local magnetic and structural properties. NMR is expected to detect any role of the magnetic fluctuation on the SPT. It may be mentioned that neither the signature of the SC transition nor SPT was detected from $^{19}$F nuclear spin-lattice relaxation rate ($1/T_1$) in SmFeAsO$_{0.85}$F$_{0.15}$ \cite{Prando10}.

\section{Experimental}
\begin{figure}
{\includegraphics{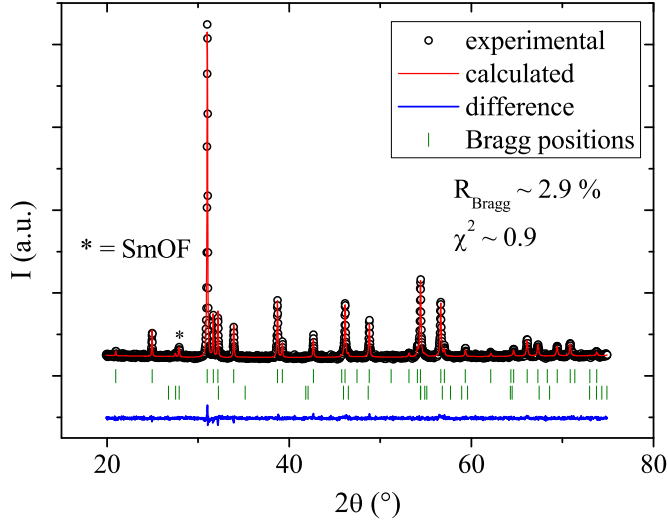}}\par \caption{(Color
online). XRD patterns of SmFeAsO$_{0.86}$F$_{0.14}$. Presence of faint amount of SmOF is marked by $\ast$. ${\chi}^2 = \displaystyle\sum_{i=1}^n w_i\{y_i - y_{ci}(\mathbf{\alpha})\}^2$ that is minimized in the Rietveld method \cite{Carvajal}, where $y_i$ = profile intensity, $y_{ci}$ = calculated counts with $\mathbf{\alpha}$ as the parameter vector and $w_i = \frac{1}{\sigma_i^2}$, $\sigma_i^2$ being the variance of the $y_i$. The quality of the agreement between the observed and calculated profiles is measured by profile factor $R_{\rm{Bragg}}$ and is defined as $R_{\rm{Bragg}} = 100\frac{\displaystyle\sum_{i=1, n}|y_i - y_{ci}|}{\displaystyle\sum_{i=1, n}y_i }$.} \label{xray}
\end{figure}

The polycrystalline sample with nominal composition SmFeAsO$_{0.86}$F$_{0.14}$ was prepared through a solid state reaction route using Sm, As, FeF$_2$, Fe and
Fe$_2$O$_3$ as starting materials. All handlings were made in an Ar-filled glove box with less than 1 ppm O$_2$ and H$_2$O. SmAs alloy was first obtained
by heating Sm and As under pure Ar in a closed silica tube at 900 $^\circ$C during 12 h. The single phase nature of the alloy was confirmed by X-ray
diffraction. SmAs was then carefully ground and sieved to less than 100 $\mu$m, and mixed in stoichiometric ratio with Fe, Fe$_2$O$_3$ and FeF$_2$, and the
resulting powder was pressed into 2$\times$3$\times$12 mm$^3$ bars under 200 MPa. These bars were heated two times with intermediate grinding and pressing, at
1150 $^\circ$C during 48 h under argon in alumina crucibles sealed in closed silica tubes, with Ta pieces as getter. XRD patterns (figure \ref{xray}) were recorded using a Philips X'Pert Pro diffractometer with X'Celerator detector using CuK$\alpha$1 radiation in a 2$\theta$ range 20 - 80$^{\circ}$. The XRD patterns were analyzed using the Rietveld method with the help of the FullProf software \cite{Carvajal}.

%%%%%%%%%%%%%%%%%%%%%%%%%%%%%%%%%%%%%%%%%%%%%%%%%%%%%%%%%%%%%%%%%
%%%%%%%%%%%%%%%%%%%%%%%%%%%%%%%%%%%%%%%%%%%%%%%%%%%%%%%%%%%%%%%%

It is very hard to determine the actual F content of 1111 compounds precisely, as F and O can not be distinguished using XRD and as the quantification of these light elements by EDX is challenging. However, we can say that the $T_c$ is consistent \cite{sanna} with a doping level of 0.14 so do the lattice parameters, obtained from Rietveld refinement, namely $a$=3.9289(5){\AA} and $c$=8.467(1){\AA} which are also consistent with this doping level as compared to the literature values \cite{Margadonna09}.

The resistivity was measured using four probe method by applying magnetic fields in the range 0-8 T using a cryogen free magnet from M/S Cryogenics Limited. The temperature variation study was performed using a close cycle refrigerator also from M/S Cryogenics Limited. The DC magnetization measurements were done in the range 0.1-7 T in PPMS of M/S Quantum Design, Inc., USA.

The $^{19}$F NMR measurements were performed on the powder sample in a fluorine free probe and carried out using a conventional phase-coherent spectrometer (Thamway PROT 4103MR) with $H_0$= 7.04 T superconducting magnet (Bruker). The spectrum was recorded by changing the frequency step by step and recording the spin echo intensity by applying a $\pi$/2-$\tau$-$\pi$/2 solid echo pulse sequence. The spin-lattice relaxation time ($T_1$) was measured using the saturation recovery method, applying a single $\pi$/2 pulse. The spin-spin relaxation time ($T_2$) was measured applying $\pi$/2-$\tau$-$\pi$ pulse sequence.

\section{Results and discussion}

\subsection{Resistivity and magnetic susceptibility}

\begin{figure}
{\includegraphics{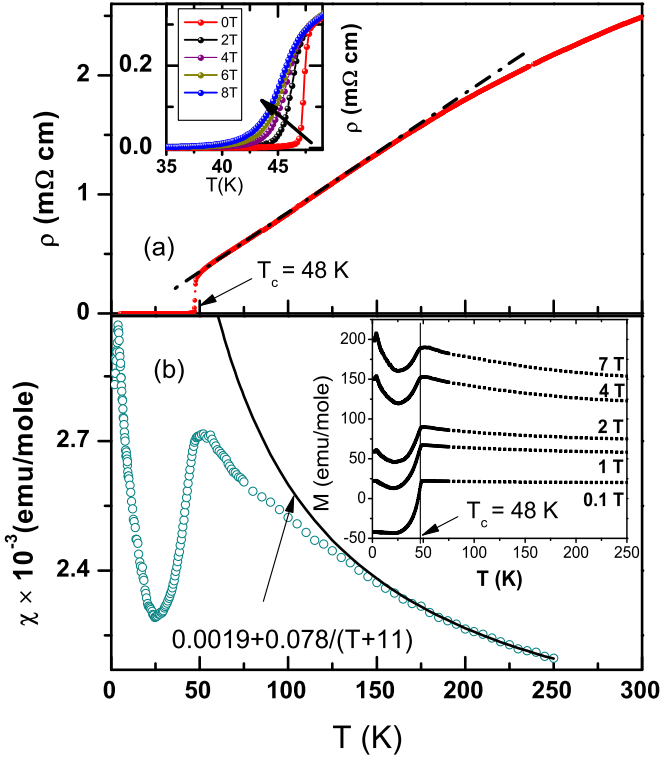}}
\caption{(Color
online). Temperature dependence of resistivity ($\rho$) and susceptibility ($\chi$) in SmFeAsO$_{0.86}$F$_{0.14}$. (a): $\rho (T)$ plot at zero field showing a sharp drop at $T_c$= 48 K, \chain: linear fit. Inset shows $\rho$ at different magnetic field. (b): $\chi(T)$ at 7 T and the solid line is the Curie-Weiss fit. The inset shows magnetization ($M$) at different field.} \label{resistivity}
\end{figure}

Figure \ref{resistivity} shows the $T$ dependence of resistivity ($\rho$) and the dc magnetic susceptibility ($\chi$). $\rho (T)$ at zero field shows a clear signature of SC transition ($T_c$) at 48 K. The width of the transition ($\Delta T_c$) =$\pm$0.8 K indicates the high
homogeneity of the sample. Inset of figure \ref{resistivity}a shows $\rho (T)$ at different magnetic field. As expected, $\Delta T_c$ is not as sharp as in zero field case.
$\rho$, above $T_c$, does not vary with $T^2$, usually seen in Fermi liquid case, instead it follows a linear behavior between $T_c$ and 160 K. Hess \etal \cite{Hess09} ascribed
the linear behavior of $\rho (T)$ above $T_c$ observed in SmFeAsO$_{1-x}$F$_{x}$ with $x$ up to 0.1, as a strange metallic behavior.
%%%%%%%%%%%%%%%%%%%%%%%%%%%%%%%%%%%%%%%%%%%%%%%%%%%%%%%%%%%%%%%%%
%%%%%%%%%%%%%%%%%%%%%%%%%%%%%%%%%%%%%%%%%%%%%%%%%%%%%%%%%%%%%%%%%%%%

Figure \ref{resistivity}(b) shows variation of $\chi$ with $T$ at 7 T with inset showing $T$ dependence of $M$ at different magnetic field. Well defined SC transition was observed at 48 K, even at 7 T. At $H$ $\geq$ 1 T, $M$ is positive below $T_c$. This is because with the increase of field, the paramagnetic moment of Sm 4$f$ (and/or Fe 3$d$) overcomes the diamagnetic response of the SC electrons.

$\chi$ - $T$ curve, in the range 160 - 300 K, follows $\chi= \chi_0 + C/(T + \theta)$, with $\chi_0$ = 0.0019 emu/mole and $\theta$ = -11 K. Where $\chi_0$ contains the contributions from Pauli-paramagnetism, orbital paramagnetism and Landau diamagnetism of conduction electrons. From the Curie-Wiess (CW) constant ($C$), we found $P_{eff}$ = 0.79 $\mu_B$, which is close to the free ion value (0.84$\mu_B$) of Sm$^{3+}$ 4$f$ local moment. The deviation of $\chi$($T$) from CW law may arise due to the development of short range magnetic correlation within Fe-3$d$ spins. Contribution from Sm-4$f$ is ruled out, as Sm 4$f$ AFM ordering temperature being $T_N$ =3.8 K (manifested as a peak in the $\chi$ - $T$ curve, shown in figure \ref{resistivity}(b)), its contribution to $\chi$ is expected to follow the CW law near 160 K. This deviation may also result from the coupling between the Fe 3$d$ spins/orbitals and the SPT as observed by Martinelli  \etal \cite{Martinelli11}. It is to be noted that the only secondary phase we have observed in the sample is from faint amount of SmOF, which is paramagnetic in the whole temperature range (see figure \ref{xray}).

\begin{figure}
\includegraphics[width=1\textwidth]{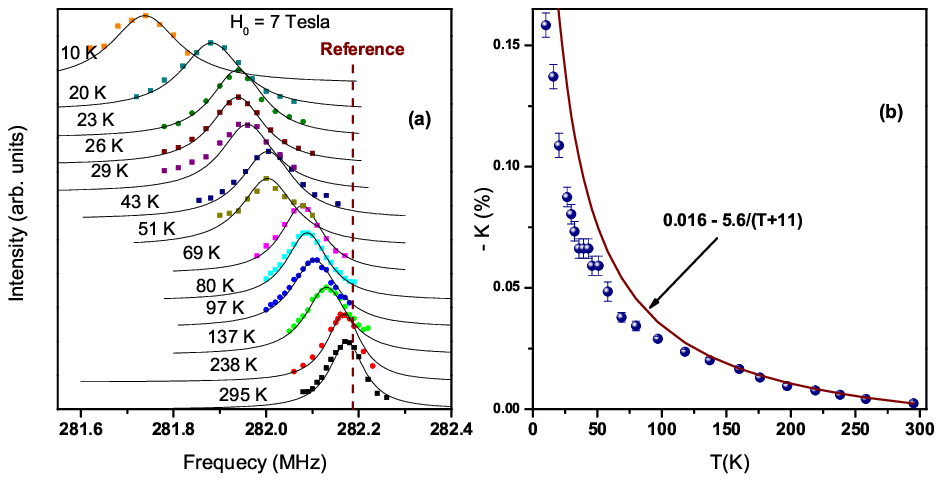}\caption{(Color
online). (a): Frequency swept $^{19}$F NMR spectra of SmFeAsO$_{0.86}$F$_{0.14}$ taken at 7.04 T. The continuous line represents the calculated spectrum. The vertical dashed line represents the resonance line position in a diamagnetic reference compound. (b): Temperature dependence of Knight shift ($K$) for $^{19}$F NMR line in SmFeAsO$_{0.86}$F$_{0.14}$. Variation of $K(T)$ showing deviation from CW fit (\full) below 160 K.} \label{Spectra-shift}
\end{figure}
\subsection{Nuclear magnetic resonance}

Figure \ref{Spectra-shift}(a) shows $^{19}$F NMR spectra along with the calculated one at different temperatures. Below 10 K, the spectra could not be detected because of the excessive line broadening/shortening of the nuclear spin-spin and spin-lattice relaxation times, due to the development of short range correlations among the Sm-4$f$ spins as revealed in the magnetic susceptibility behavior. Almost symmetric (lorentzian) spectrum indicates negligible anisotropy in the local magnetic field at the $^{19}$F site as was also reported in (La/Sm)FeAsO$_{1-x}$F$_{x}$ \cite{Ahilan08,Prando10}. The shift ($K$) of the resonance line (from the diamagnetic reference position) was measured either from the peak position of the spectra or obtained from the theoretically fitted curve and is negative in sign (figure \ref{Spectra-shift}(b)) throughout the whole temperature range. In LaFeAsO$_{0.9}$F$_{0.1}$, $^{19}$F NMR showed small positive shift $\sim$ 100 ppm above $T_c$ \cite{Ahilan08}. This indicates that in SmFeAsO$_{0.84}$F$_{0.14}$, the dominant contribution to the Knight shift arises from the Sm-4$f$ spins over that of Fe-3$d$. Moreover,
 in case of SmCoPO, where phosphorous lies in the Co-P plane, it was shown from $^{31}$P NMR \cite{Majumder12} that the major contribution to the Knight shift arises from Sm-4$f$ electrons.

The $T$ dependence of the shift, $K$ is given by
\begin{equation}
K = K_0 + (H_{hf}/N\mu_B)\chi(T)
\end{equation}
where $K_0$ is the $T$ independent contribution arising from Pauli-paramagnetic and orbital parts of the conduction electrons, and transferred Van-Vleck susceptibility from rare-earth ion. $H_{hf}$ represents the hyperfine coupling constant between the $^{19}$F nuclear spin and the Fe-3$d$ and Sm-4$f$ electron spins, $N$ and $\mu_B$ are the Avogadro number and Bohr magneton respectively. $K$ - $T$ curve
in the range 160 - 300 K, fitted (figure \ref{Spectra-shift}(b): solid line) using the CW
formula for $\chi(T)$ gives $K_0$= 0.016\%, $\theta$ = -11 K, and the product of the $C$ and
$H_{hf}$. Using $C$ from $\chi - T$ curve, we have $H_{hf}$ = -3.93$\pm$0.06 kOe/$\mu_B$. The negative hyperfine coupling arises due to the dominant contribution of Sm-4$f$ over that of Fe-3$d$, as the $^{19}$F $H_{hf}$ is positive in F- doped LaFeAsO \cite{Ahilan08}. In this connection it may be mentioned that the $^{19}$F Knight shift data is not available in $F$-doped PrFeAsO.

\subsection{Nuclear spin-spin ($1/T_2$) and spin-lattice $1/T_1$ relaxation rates}

\begin{figure}
{\centering {\includegraphics[width=.6\textwidth]{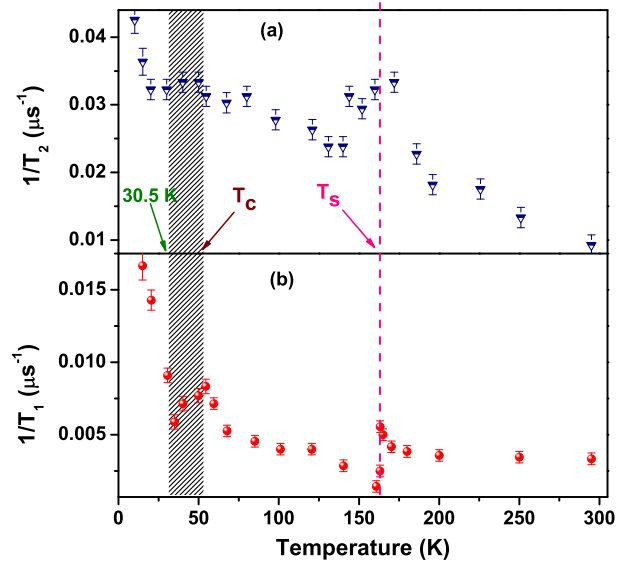}}\par} \caption{(Color
online). Temperature dependence of $^{19}$F NMR (a): spin-spin $1/T_2$ and (b): spin-lattice $1/T_1$ relaxation rate in SmFeAsO$_{0.86}$O$_{0.14}$.}\label{relaxation}
\end{figure}

Figure \ref{relaxation} shows $^{19}$F NMR transverse relaxation rate, $1/T_2$ (gives the intrinsic width of the resonance line) and the longitudinal relaxation rate, $1/T_1$ as a function
of $T$. In the temperature range 10 - 300 K, except between a narrow range 160 - 170 K, $T_1$ was determined from the recovery of the longitudinal component of the nuclear magnetization $M(\tau$) as a function of the delay
time $\tau$ using $M(\tau) = M(\infty)(1-\exp(-\tau/T_1))$. In the latter range we have used
a stretched exponential ($M(\tau) = M(\infty)(1-\exp(-\tau/T_1)^\beta)$) to fit the recovery curve with the exponent $\beta$=0.75 (figure \ref{recovery}).
The recovery of the transverse magnetization was found to be a single exponential at all temperatures. $T_2$ was obtained by fitting the decay curve with the equation $M(2\tau) = M_0\exp(-2\tau/T_2$), where $M_0$ is the initial magnetization.

%%%%%%%%%%%%%%%%%%%%%%%%%%%%%%%%%%%%%%%%%%%%%%%%%%%%%%%%%%%%%%%%%
%%%%%%%%%%%%%%%%%%%%%%%%%%%%%%%%%%%%%%%%%%%%%%%%%%%%%%%%%%%%%%%%%%%%

Above 180 K, $1/T_1$ remains constant while $1/T_2$ shows continuous increment. Values of $1/T_1$ are almost same as reported by Prando \etal \cite{Prando10} and three orders of magnitude larger than in LaFeAsO$_{1-x}$F$_{x}$ \cite{Ahilan08}. This implies dominant contribution of localized Sm 4$f$ spin-fluctuation on the $1/T_1$ and $1/T_2$ processes \cite{Moriya85,Guguchia11}. At around 163 K, both $1/T_1$ and $1/T_2$ data show a clear peak. This we associate to the SPT as reported by Martinelli \etal \cite{Martinelli11}. We believe this observation is the first evidence of SPT in SmFeAsO$_{0.86}$O$_{0.14}$ using any microscopic tool. We shall discuss the nature of this SPT in the following section.

Above 180 K, $1/T_1$ remains constant while $1/T_2$ shows continuous increment. Values of $1/T_1$ are almost same as reported by Prando \etal \cite{Prando10} and three orders of magnitude larger than in LaFeAsO$_{1-x}$F$_{x}$ \cite{Ahilan08}. This implies dominant contribution of localized Sm 4$f$ spin-fluctuation on the $1/T_1$ and $1/T_2$ processes \cite{Moriya85,Guguchia11}. At around 163 K, both $1/T_1$ and $1/T_2$ data show a clear peak. This we associate to the SPT as reported by Martinelli \etal \cite{Martinelli11}. We believe this observation is the first evidence of SPT in SmFeAsO$_{0.86}$O$_{0.14}$ using any microscopic tool. We shall discuss the nature of this SPT in section \ref{SPT}.

Below SPT down to 50 K, both $1/T_1$ and $1/T_2$ show slow enhancement. However, in the range 30 K $< T < T_c$, $1/T_1$ shows a small but clear decrement which may be associated with the enhanced contribution from 4$f$ local spin ordering superimposed on the decreasing trend of the diamagnetic contribution of the superconducting electrons. It is to be mentioned that the magnitude of $^{19}$F NMR $1/T_1$ in the normal state of PrFeAsO$_{0.89}$F$_{0.11}$ is very close to that in LaFeAsO$_{0.89}$F$_{0.11}$, indicating a negligible contribution of Pr-4$f$ spin over Fe 3$d$ \cite{Matano08}. Moreover, the effect of Pr-4$f$ spin ordering on the $T$ dependence of $1/T_1$ (as well as in Knight shift) was not observed. This makes it possible to observe a large drop in $1/T_1$ at $T_c$ in PrFeAsO$_{0.89}$F$_{0.11}$.

Finally, the sharp increase of 1/$T_1$ and $1/T_2$ in the range 10 - 25 K should be due to the development of Sm-4$f$ spin short range correlation (which was also reflected in the behavior of the bulk susceptibility), as the system approaches $T_N$ (3.8 K) \cite{Ryan09}.

\subsection{The nature of the structural transition}\label{SPT}
\begin{figure}
\includegraphics[width=1\textwidth]{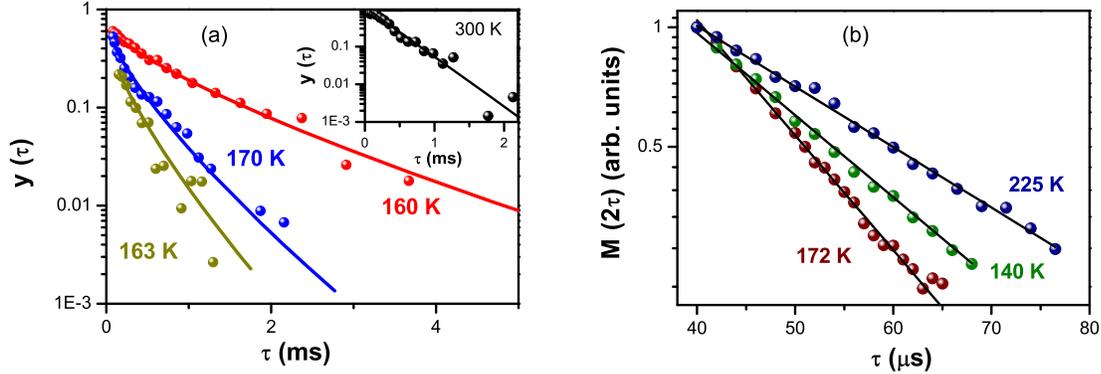}\caption{(Color
online). Recovery curves for (a): longitudinal ($T_1$), ($y(\tau) = \frac{M(\infty)- M(\tau)}{M(\infty)})$ and (b): transverse ($T_2$) relaxation times of SmFeAsO$_{0.86}$F$_{0.14}$.} \label{recovery}
\end{figure}

We have shown the existence of anamoly (due to SPT?) near 160 K in the temperature variation of resistivity, susceptibility and the $^{19}$F NMR Knight shift in SmFeAsO$_{0.86}$O$_{0.14}$. However, the most direct evidence of SPT has come from $1/T_1$ and $1/T_2$ measurement.
Occurrence of SPT will be reflected from $^{19}$F ($I$ = $\frac{1}{2}$) NMR $1/T_1$ through a change in the electron-nuclear dipolar and hyperfine contribution to the fluctuating local magnetic field ($H_{local}$) produced at the $^{19}$F site by the Fe-3$d$ and Sm-4$f$ spin. On the other hand $1/T_2$ is governed by the contributions from (i) dipolar interaction of the $^{19}$F nuclear spins with the like and unlike nuclear spins, which is temperature and field independent (1/T$_2|_{static}$) and (ii) the dipolar and hyperfine interactions of the $^{19}$F nuclei with the longitudinal component of the fluctuating magnetic field produced by the neighboring Fe$^{2+}$ 3$d$-spins and Sm 4$f$-spins (1/T$_2|_{dynamic}$). For localized spin, the latter contribution shows a Curie-Weiss behavior in $T$ dependence \cite{Guguchia11}. Close to the magnetic ordering temperature, it shows strong $T$ dependence, as the fluctuation frequency of $H_{local}$ reduces and becomes comparable to the nuclear resonance frequency due to the development of short range magnetic correlation. Near SPT ($T_S$), as the atomic positions are altered, it should change both (1/T$_2|_{static}$) and 1/T$_2|_{dynamic}$.
The magnitude of the drop in both $1/T_1$ and $1/T_2$ (clearly revealed from the change in the nature of the decay curves (figure \ref{recovery})) below $T_S$ is small indicating a weak tetragonal to orthorhombic distortion as also pointed out by Martinelli et al from structural study. Nevertheless, the effect of this small orthorhombic distortion is reflected in both $1/T_1$ and $1/T_2$ data. As pointed out earlier, the $T$ independent behavior of $1/T_1$ in the range 180 - 300 K indicates faster fluctuations of $H_{local}$ than the NMR resonance frequency ($\nu_R$). The enhancement of $1/T_1$ in the range 163-175 K is a clear signature of the slowing down of the fluctuation of $H_{local}$, so that it's frequency is comparable to $\nu_R$, contributing to the relaxation process. Such slowing down of the fluctuation of $H_{local}$ can arise due to several reasons. (1) It may be due to the development of a short range magnetic correlations of Fe-3$d$ spins (as the Sm-4$f$ spins order magnetically at 3.8 K ($T_N$), the short range correlation among the 4$f$ spins would occur close to $T_N$, so its temperature dependent contribution to $1/T_1$ would arise far below 175 K). This is supported by the fact that $\chi$ and $K$ vs $T$ curve deviate from CW law below 160 K (figure \ref{resistivity}(b) and \ref{Spectra-shift}(b)) with $\theta$ = -11 K implying AFM correlations. Thus a role of AFM correlated spin fluctuations (may be secondary) in driving the SPT can not be ruled out. (2) The effect of slowing down of the orbital fluctuations near $T_S$ may also have a contribution (through spin-orbit coupling), if SPT is driven by 3$d$ orbital ordering as proposed theoretically \cite{Weicheng09} and from synchrotron powder diffraction data \cite{Martinelli11}. (3) Softening of a lattice vibrational mode near $T_S$ can also enhance $1/T_1$ provided it can reduce the fluctuation frequency of $H_{local}$ through a coupling with the spin /orbital motions of the electrons, as the contribution of phonon modes to $1/T_1$ is much less compared to that of the fluctuations of $H_{local}$.

We have shown that within a short interval 170 - 160 K, the recovery curves of $T_1$ (figure \ref{recovery}(a)) follow a stretched exponential $M(\tau) = M(\infty)(1 - \exp(-\tau/T_1)^\beta)$ with the exponent $\beta$ $\sim$ 0.75$\pm$0.03 indicating a distribution of $1/T_1$. Spatial distribution of the SPT temperature can induce a stretched exponential in the recovery curve but the clear sharp peak in $1/T_1$ (figure \ref{relaxation}(b)) eliminates this possibility. However, the SPT related disorder may induce a distribution of $T_1$ over a certain temperature range above $T_S$ resulting in the necessity of a stretched exponential. At the final stage of preparation of this manuscript, we have noticed the work of Martinelli \etal \cite{Martinelli12}, where they have shown that on cooling, micro strain along the tetragonal \emph{hh}0 direction appears and increases as the temperature is decreased. Just above the structural transition, micro strain reaches its maximum value and then is abruptly suppressed by symmetry breaking. Micro strain reflects a distribution of the lattice parameters in the tetragonal phase and hence explain the occurrence of stretched exponential in the recovery curves of $T_1$.
From NMR results it is therefore confirmed microscopically that in optimally F doped SmFeAsO system which exhibit superconductivity, the tetragonal to orthorhombic transition near 163 K is still present, as in the non superconducting parent compound. This finding creates interest to explore whether this occurs only in Sm based Fe-As systems or also in other rare-earth based 1111 superconductors.

\section{Conclusion}
In conclusion, we report the resistivity, magnetization and $^{19}$F NMR results in superconducting SmFeAsO$_{1-x}$F$_{14}$. Both the resistivity and the magnetization data clearly show the signature of the superconducting transition at 48 K. The behavior of 1/$T_1$ and 1/$T_2$ confirms, for the first time from a microscopic tool like NMR, the existence of a structural phase transition at 163 K, as in the non superconducting parent compound, in a sample which undergoes superconducting transition at 48 K. This finding creates interest to explore whether this is unique for only Sm based  systems or it is also present in other rare-earth based 1111 superconductors.

\ack
We sincerely thank Prof. Takao Kohara of University of Hyogo, Japan for discussion and useful suggestions and Mr. D.J. Seth for his help in constructing a fluorine free probe for NMR measurement.

\end{document}